\def\rfr#1{Equation (\ref{#1})}

\def\virg#1{``#1''}

\def\eqi{\begin{equation}}
\def\eqf{\end{equation}}
\def\eqia{\begin{eqnarray}}
\def\eqfa{\end{eqnarray}}

\def\rp#1#2{{#1\over#2}}
\def\lb#1{\label{#1}}

\def\bds#1{\boldsymbol{#1}}

\def\ton#1{\left(#1\right)}
\def\qua#1{\left[#1\right]}

\documentclass[Universe,article,accept,oneauthor,12pt,a4paper]{mdpi}

\lastpage{82}
\doinum{10.3390/universe1010038}
\pubvolume{1}
\pubyear{2015}
\history{\vspace{-12pt}Received: 6 March 2015  / Accepted: 17 April 2015 / Published: 24 April 2015}

\usepackage{amsmath,amssymb,amsthm,amscd,latexsym,textgreek,w-greek}
\usepackage[polutonikogreek,english]{babel}
\usepackage{isomath}
\usepackage{hyperref}
\usepackage[combine,document]{ucs}
\usepackage[utf8x]{inputenx}
\usepackage[LGR,T1]{fontenc}
\usepackage[toc,title,titletoc]{appendix}
\usepackage{txfonts}
\usepackage{graphicx,epsfig}
\RequirePackage{color}
\allowdisplaybreaks 
\newcommand{\grk}[1]{\selectlanguage{polutonikogreek}
#1\selectlanguage{english}}

\usepackage{enumitem}
\setitemize{parsep=0pt,itemsep=0pt,leftmargin=.4in}
\setenumerate{parsep=0pt,itemsep=0pt,leftmargin=.4in}

\Title{Editorial for the Special Issue 100 Years of Chronogeometrodynamics: The Status of the Einstein's \protect\linebreak Theory of Gravitation in Its Centennial Year}

\Author{Lorenzo Iorio}

\address[1]{Ministero dell' Istruzione dell' Universit\`{a} e della Ricerca (M.I.U.R), Fellow of the Royal Astronomical Society (F.R.A.S.), Viale Unit\`{a} di Italia, 68 70125 Bari, Italy; E-Mail: lorenzo.iorio@libero.it; \mbox{Tel.: +39-3292399167}
\vspace{-12pt}
}

\abstract{The present Editorial introduces the Special Issue dedicated by the journal {\em Universe} to the General Theory of Relativity, the beautiful theory of gravitation of Einstein, a century after its birth. It reviews some of its key features in a historical perspective, and, in welcoming distinguished researchers from all over the world to contribute it, some of the main topics at the forefront of the current research are outlined.
}

\keyword{general relativity and gravitation; classical general relativity; gravitational waves; quantum gravity; cosmology; experimental studies of gravity}

\PACS{04.; 04.20.-q; 04.30.-w; 04.60.-m; 98.80.-k; 04.80.-y}

\begin{document}

\vspace{-12pt}
\section{Introduction}\lb{inizio}

This year marks the centenary of the publication of the seminal papers \cite{Ein15a, Ein15b, Ein15d} in which Albert Einstein laid down the foundations of his theory of gravitation, one of the grandest achievements of the human thought which is the best description currently at our disposal of such a  fundamental interaction shaping the fabric of the natural world. It is usually termed \virg{General Theory of Relativity} (GTR, from {\em Allgemeine Relativit\"{a}tstheorie}), often abbreviated as \virg{General Relativity} (GR). It replaced the Newtonian concept of \virg{gravitational force} with the notion of \virg{deformation of the chronogeometric structure of spacetime} \cite{dam0} due to all forms of energy weighing it; as such, it can be defined as a chronogeometrodynamic theory of gravitation \cite{torr0}.

 GTR is connected, in a well specific sense, to another creature of Einstein himself, with Lorentz \cite{lor1} and Poincar\'{e} \cite{poin1, poin2}  as notable predecessors, published in 1905 \cite{Ein05}: the so-called Special (or Restricted) Theory of Relativity (STR). The latter is a physical theory whose cornerstone is the requirement of covariance of the differential equations expressing the laws of physics (originally only mechanics and electromagnetism) under Lorentz transformations of the spacetime coordinates connecting different inertial reference frames, in each of which they must retain the same mathematical form. More precisely,~if \eqi A\ton{x,y,z,t},B\ton{x,y,z,t},C\ton{x,y,z,t},\ldots\eqf represent the state variables of a given theory depending on spacetime coordinates $x,y,z,t$ and {are} mutually connected by some mathematical  relations \eqi f(A,B,C,\ldots)=0\eqf representing the theory's fundamental equations, the latter ones can always be mathematically written in a covariant form under a generic transformation from the old coordinates to the new ones  as \eqi f^{'}\left(A^{'}\ton{x^{'},y^{'},z^{'},t^{'}},B^{'}\ton{x^{'},y^{'},z^{'},t^{'}},C^{'}\ton{x^{'},y^{'},z^{'},t^{'}},\ldots\right) = 0 \eqf In general, the new functional relations $f^{'}$ connecting the transformed state variables $A^{'},B^{'},C^{'},\ldots$ are different from the {ones of} $f$. If, as for the Lorentz transformations, it turns out \eqi f^{'} = f \eqf which does not necessarily implies that also the state variables $A,B,C,\ldots$ remain unchanged, then it is said that the equations of the theory retain the same form. It is just the case of the Maxwell equations, in which the electric and magnetic fields $\bds E,~\bds B$ transform in a given way under a Lorentz transformation in order to keep the form of the equations connecting them identical, which, instead, is not retained under Galilean transformations \cite{invconv}. In the limiting case of the {Galilean} transformations applied to the Newtonian mechanics, it turns out that the theory's equations are even invariant in the sense that also the state variables remain unchanged, \emph{i.e}., it is \eqi{\bds F}^{'} -m {\bds a}^{'} = 0\eqf with
 \begin{align}
 {\bds F}^{'} &= \bds F\\ \nonumber
 {\bds a}^{'} &= \bds a
 \end{align}

 As such, strictly speaking, the key message of STR is actually far from being: \virg{everything is relative}, as it might be seemingly suggested by its rather unfortunate name which, proposed for the first time by Planck \cite{Planck} ({\em Relativtheorie}) and Bucherer \cite{Buch06} ({\em Relativit\"{a}tstheorie}), became soon overwhelmingly popular (see also \cite{CPAE89}). Suffice it to say that, in informal correspondence,  Einstein himself  would have preferred that its creature was named as {\em Invariantentheorie} (Theory of invariants) \cite{Hol}, as also explicitly proposed-unsuccessfully-by Klein \cite{Klein}. Note that, here, the adjective \virg{invariant} is used, in a looser sense, to indicate the identity of the mathematical functional form connecting the transformed \mbox{state~variables}.

Notably, if the term \virg{relativity} is, instead, meant as the identity of all physical processes in reference frames in reciprocal translational uniform motion connected by Lorentz transformations, then, as remarked by Fock \cite{Focka}, a name such as \virg{Theory of Relativity} can, to some extent, be justified. In {\em this specific sense}, relativity geometrically corresponds to the maximal uniformity of the pseudo-Euclidean spacetime of Poincar\'{e} and Minkowski in which it is formulated. Indeed, given a $N-$dimensional manifold, which can have constant curvature, or, if with zero curvature, can be Euclidean or pseudo-Euclidean, the group of transformations  which leave identical the expression for the squared distance between two nearby points can contain at most $\ton{1/2}N(N+1)$ parameters. If there is a group involving all the $\ton{1/2}N(N+1)$ parameters, then the manifold is said to have maximal uniformity. The most general Lorentz transformations, which leave unchanged the coefficients of the expression of the 4-dimensional distance between two nearby spacetime events, involve just 10 parameters. Now, in the pseudo-Riemannian spacetime of GTR the situation is different because, in general, it is not uniform at all in the geometric sense previously discussed. Following Fock \cite{Focka}, it can be effectively  illustrated by a simple example whose conclusion remains valid also for the geometry of the $4$-dimensional spacetime manifold. Let us think about the surface  of a sphere, which is a 2-dimensional manifold of a very particular form. It is maximally uniform since it can be transformed into itself by means of rotations by any angle about {an arbitrary} axis passing through the centre, so that the associated group of transformations has just three parameters. As a result, on a surface of a sphere there are neither preferred points nor preferred directions. A more general non-spherical surface of revolution has only partial uniformity since it can be transformed into itself by rotation about an axis which is now fixed, so that the rotation angle is the only arbitrary parameter left. There are privileged points and lines: the poles through which the axis passes, meridians, and latitude circles. Finally, if we consider a  surface of general form, there will be no transformations taking it into itself, and it will possess no uniformity whatsoever. Thus, it should be clear that the generality of the form of the surface is a concept antagonistic to the concept of uniformity. Returning now to the concept of relativity {\em in the aforementioned specified sense}, it is related to uniformity in all those cases in which the spacetime metric can be considered fixed. This occurs not only in the Minkowskian spacetime, but also in the Einsteinian one, provided only that the physical processes one considers have no practical influence on the metric. Otherwise, it turns out that relativity can, to a certain extent, still be retained only if the non-uniformity generated by heavy masses may be treated as a local perturbation in infinite Minkowskian spacetime. To this aim, let us think about a laboratory on the Earth's surface \cite{Focka}. If it was turned upside down, relativity would be lost since the physical processes in it would be altered. But, if the upset down laboratory was also parallel transported to the antipodes, relativity would be restored since the course of all the processes would be the same as at the beginning. In this example, a certain degree of relativity was preserved, even in a non-uniform spacetime, because the transformed gravitational field $g^{'}$ in the new coordinate system $\left\{x^{'}\right\}$ has the same form as the old field $g$ in the old coordinates $\left\{x\right\}$, \emph{i.e}.,

 \vspace{-12pt}
 \begin{align}
 \left\{x\right\}&\mapsto \left\{x^{'}\right\}\\
 \vspace{-12pt}g\left(x\right)&\mapsto g^{'}\left(x^{'}\right) = g\left(x\right)
 \end{align}
 \vspace{-12pt}

 Such considerations should have clarified that relativity, in the previously specified sense, either does no exist at all in a non-uniform spacetime like the Einsteinian one, or else it does exist, but does {\em not} go {\em beyond} the relativity of the Minkowskian spacetime. {\em In this sense}, the gravitational theory of Einstein {\em cannot} be a {\em generalization of his theory of space and time of 1905}, and its notion of relativity along with its related concept of maximal uniformity was {\em not} among the concepts  subjected to {\em generalization}. Since the greatest possible uniformity is expressed by Lorentz transformations, there {\em cannot} be a {\em more general} principle of relativity than that discussed in the theory of 1905. All the more, there cannot be a general principle of relativity having physical meaning which would hold with respect to arbitrary frames of references. As such, both the denominations of \virg{General Relativity} and \virg{General Theory of Relativity}
are confusing and lead to misunderstandings.  Furthermore, such adjectives reflect also an incorrect understanding of the theory itself since they were adopted referring to the covariance of the equations with respect to arbitrary transformations of coordinates accompanied by the transformations of the coefficients of the distance between two events in the $4$-dimensional spacetime. But it turned out that such kind of covariance has actually nothing to do with the uniformity or non-uniformity of spacetime \cite{Cartan, Focka}. Covariance of equations {\em per se} is just a merely mathematical property which in no way is expression of any kind of physical law. Suffice it to think about the Newtonian mechanics and the physically equivalent Lagrange equations of second kind which are covariant with respect to arbitrary transformations of the coordinates. Certainly, nobody would state that Newtonian mechanics contains in itself \virg{general} relativity. A principle of relativity-Galilean or Einsteinian-implies a covariance of equations, but the converse  is not true: covariance of differential equations is possible also when no principle of relativity is satisfied. Incidentally, also the the adjective \virg{Special} attached to the theory of 1905 seems improper in that it purports to indicate that it is a special case of \virg{General} Relativity.

In the following, for the sake of readability, we will adhere to the time-honored conventions by using STR and GTR (or GR) for the Einsteinian theory of space and time of 1905 and for his gravitational theory of 1915, respectively.

Of course, the previous somewhat \virg{philosophical} considerations are, by no means, intended to undermine the credibility and the reliability of the majestic theory of gravitation by Einstein, whose concordance with experiments and observations has been growing more and more over the latest decades~\cite{WillLRR}.

Below, some key features of GTR, to which the present Special Issue is meritoriously and timely dedicated, are resumed in a historical perspective \cite{Pau21, torr, Renn}, without any pretence of completeness. It is hoped that the distinguished researchers  who will kindly want to contribute it will provide the community of interested readers with the latest developments at the forefront of the research in this fascinating and never stagnant field.

{In the following, Greek letters  $\mu,\nu,\varrho\ldots$ denote 4-dimensional spacetime indexes running over $0,1,2,3$, while Latin ones $i,j,k,\ldots$, taking the values $1,2,3$, are for the 3-dimensional space.}
\section{The Incompatibility of the Newtonian Theory of Gravitation with STR}\lb{incompatibilita}

In the framework of the Newtonian theory of universal gravitation \cite{princi}, the venerable force-law yielding the acceleration $\bds a$ imparted on a test particle by a mass distribution of density $\rho$ could be formally reformulated in the language of the differential equations governing a field-type state variable $\Phi$, known as potential, through the Poisson equation \cite{poisson} \eqi{\bds \nabla}^2 \Phi = 4\pi G \rho\eqf where $G$ is the Newtonian constant of gravitation, so that \eqi\bds a = -{\bds\nabla}\Phi\eqf

Nonetheless, although useful from a mathematical point of view, such a field was just a non-dynamical entity, deprived of any physical autonomous meaning: it was just a different,  mathematical way of telling the same thing as the force law actually did \cite{torr}. It is so because, retrospectively, in the light of STR,  it was as if, in the Newtonian picture,  the gravitational interaction among bodies would take place {\em de facto} instantaneously, irrespectively of the actual distance separating them, or as if gravity would be some sort of occult, intrinsic property of matter itself. Remarkably, such a conception was opposed by Newton himself who, in the fourth letter to R. Bentley in 1692, explicitly wrote \cite{lettera}: \virg{[\ldots] Tis inconceivable that inanimate brute matter should (without the mediation of something else which is not material) operate upon $\&$ affect other matter without mutual contact; as it must if gravitation in the sense of Epicurus be essential $\&$ inherent in it. And this is one reason why I desired you would not ascribe \{innate\} gravity to me. That gravity should be innate inherent $\&$ \{essential\} to matter so that one body may act upon another at a distance through a vacuum without the mediation of any thing else by $\&$ through which their action or force \{may\} be conveyed from one to another is to me so great an absurdity that I believe no man who has in philosophical matters any competent faculty of thinking can ever fall into it. Gravity must be caused by an agent \{acting\}  consta\{ntl\}y according to certain laws, but whether this agent be material or immaterial is a question I have left to the consideration of my readers.}. In the previous quotation, the text in curly brackets \{$\ldots$\} is unclear in the manuscript, but the editor of the original document is highly confident of the reading.

In the second half of the nineteenth century, with the advent of the Maxwellian field theory of electromagnetism \cite{Maxw} scientists had at disposal a mathematically coherent and empirically well tested model of a physical interaction among truly dynamical fields which propagate as waves even {\em in vacuo}  at the finite speed of light $c$  transferring energy, momentum and angular momentum from a point in space to another. Now, STR is based on two postulates: The Principle of Relativity, extended by Einstein to all physical interactions, and another principle that states that the speed of light is independent of the velocity of the source. In this form, it retains its validity also in GTR. The latter is an immediate consequence of the law of propagation of an electromagnetic wave front which is straightforwardly obtained from the Maxwell equations obeying, by construction, the Principle of Relativity itself since they turned out to be covariant under Lorentz transformations. It necessarily follows \cite{Focka} that there exists a maximum speed for the propagation of any kind of physical action. This is numerically equal just to the speed of light {\em in vacuo}. If there was no single limiting velocity but instead different agents, e.g., light and gravitation, propagated {\em in vacuo} with different speeds, then the Principle of Relativity would necessarily be violated as regards at least one of the the agents. Indeed, it would be possible to choose an inertial frame traveling just at the speed of the slower agent in which the differential equations governing its course would take a particular form with respect to that assumed in all the other frames, thus predicting spurious, unphysical phenomena. It is reminiscent of the famous first {\em gedankenexperiment} made by Einstein  about STR around 1895-1896 described by himself as follows \cite{Ein56}: \virg{[\ldots] Wenn man einer Lichtwelle mit Lichtgeschwindigkeit nachl\"{a}uft, so w\"{u}rde man ein zeitunabh\"{a}ngiges Wellenfeld vor sich haben. So etwas scheint es aber doch nicht zu geben!} [\virg{If one goes after a light wave with light velocity, then one would have a time-independent wavefield in front of him. However, something like that does not seem to exist!}] Indeed, the Maxwell equations {\em in vacuo}, in their known form, do not predict stationary solutions.
That posed severe challenges to the Newtonian gravitational theory \cite{poin2}, which necessarily would have had to abandon its strict force-law aspect in favor of a genuine field-type framework making the Poisson equation covariant under Lorentz transformations \cite{Abr12, Pau21}.

Furthermore, as pointed out by Einstein himself \cite{Ein11}, Newtonian universal gravitation did not fit into the framework of the maximally uniform spacetime of SRT for the deepest reason that \cite{Focka}, while in SRT the inertial mass $m_{\rm i}$ of a material system had turned out to be dependent on its total energy, in the Newtonian picture the gravitational mass $m_{\rm g}$, did not. At high speeds, when the change in the inertia of a body becomes notable, this would imply a breakdown of the law of free fall, whose validity was actually well tested, although only at non-relativistic regimes (see Section \ref{EP}).

Finally, it can be remarked also that the required Lorentz covariance would have imposed, in principle,  also the existence of a new, magnetic-type component of the gravitational field so to yield some sort of gravitational inductive phenomena and travelling waves propagating at the finite speed of light {\em in vacuo}. Unfortunately, at the dawn of the twentieth century, there were neither experimental nor observational evidence of such postulated manifestations of a somehow relativistic theory of gravitation.

%
\section{The Equivalence Principle and Its Consequences}\lb{EP}
\vspace{-12pt}
\subsection{The Equality of the Inertial and Gravitational Masses Raised to the Status of a Fundamental Principle of Nature}

Luckily, at that time, Einstein was pressed also by another need: The quest for a coherent framework to consistently write down the laws of physics in arbitrary frames of references moving according to more complicated kinematical laws than the simple uniform translation. In 1907 \cite{jahr07}, Einstein realized that the bridge across such two apparently distinct aspects could have been represented by the equality of the inertial and gravitational masses, known  at that time to a $5\times 10^{-8}$ accuracy level thanks to the E\"{o}tv\"{o}s experiment \cite{eot}.

That was an empirical fact well known since the times of Galilei thanks to the (likely) fictional \cite{gal1, gal2, gal3} tales of his evocative free fall experiments \cite{gal4} allegedly performed from the leaning tower of Pisa around 1590. Newton himself was aware of the results by Galilei, and made his own experiments with pendulums of various materials obtaining an equality of inertial and gravitating masses to a $10^{-3}$ level of relative accuracy. Indeed, in the Proposition \textsc{VI}, Theorem \textsc{VI}, Book \textsc{III} of his {\em Principia} \cite{princi} Newton wrote \cite{chandra1}: \virg{It has been, now for a long time, observed by others, that all sorts of heavy bodies [\ldots] descend to the Earth from equal heights in equal times; and that equality of times we may distinguish  to a great accuracy, by the help of pendulums.  I tried experiments with gold, silver, lead, glass, sand, common salt, wood, water, and wheat. I provided two wooden boxes, round and equal: I filled the one with wood, and suspended an equal weight of gold (as exactly as I could) in the centre of oscillation of the other. The boxes, hanging by equal threads of 11 feet, made a couple of pendulums perfectly equal in weight and figure, and equally receiving the resistance of the air. And, placing the one by the other, I observed them to play together forwards and backwards, for a long time, with equal vibrations. And therefore the quantity of matter in the gold (by Cors. I and VI, Prop. XXIV, Book II) was to the quantity of matter in the wood as the action of the motive force (or {\em vis motrix}) upon all the gold  to the action of the same upon all the wood; that is, as the weight of the one to the weight of the other: and the like happened in the other bodies. By these experiments, in bodies of the same weight, I could manifestly have discovered a difference of matter less than the thousandth part of the whole, had any such been.} Interestingly, in the Proposition \textsc{VI}, Theorem \textsc{VI}, Book \textsc{III} of his {\em Principia} \cite{princi}, Newton looked also the known motions of the natural satellites of Jupiter to make-from a phenomenological point of view
-a further convincing case for the equality of the inertial and gravitational masses. Indeed, if the ratios of the gravitational to the inertial mass of Jupiter and of its satellites were different, the orbits of the Jovian moons about their parent planet would be unstable because of an imperfect balancing of the centrifugal acceleration and the Jupiter centripetal attraction caused by a residual, uncancelled force due to the Sun's attractions on either Jupiter and its moons themselves. Indeed, Newton wrote  \cite{chandra2}: \virg{[\ldots] that the weights of Jupiter and of his satellites towards the Sun are proportional to the several quantities of their matter, appears from the exceedingly regular motions of the satellites (by Cor. \textsc{III}, Prop. \textsc{LXV}, Book \textsc{I}).\linebreak For if some of those bodies were more strongly attracted to the Sun in proportion to their quantity of matter than others, the motions of the satellites would be disturbed by that inequality of attraction (by Cor. \textsc{II}, Prop. \textsc{LXV}, Book \textsc{I}). If, at equal distances from the Sun, any satellite, in proportion to the quantity of its matter, did gravitate towards the Sun with a force greater than Jupiter in proportion to his, according to any given proportion, suppose of $d$ to $e$; then the distance between the centres of the Sun and of the satellite's orbit would be always greater than the distance between the centres of the Sun and of Jupiter, nearly as the square root of that proportion: as by some computations I have found. [\ldots]} In principle, the Newtonian gravitational theory  would have not lost its formal consistency even if experiments-all conducted at low speeds with respect to $c$-would have returned a different verdict about $m_{\rm i}/m_{\rm g}$. Nonetheless, one cannot help but notice as the very same name chosen by Newton for the universally attractive force regulating the courses of the heavens, \emph{i.e}., gravitation,  may point, somehow, towards a not so accidental nature of the equality of inertial and gravitating masses. Indeed, it comes from the Latin word {\em gravis} (`heavy') with several Indoeuropean cognates \cite{gravis}, all with approximately the same meaning related to the weight of common objects on the Earth's surface: Sanskrit, {\em guru\d{h}} (`heavy, weighty, venerable'); Greek, \grk{b'aros} (`weight') and \grk{bar'us} (`heavy in weight'); Gothic, {\em kaurus} (`heavy'); Lettish, {\em gruts} (`heavy'). It is tempting to speculate that, perhaps, Newton had some sort of awareness of the fundamental nature of that otherwise merely accidental fact. It seems not far from the position by Chandrasekhar who wrote~\cite{chandra3}: \virg{There can be no doubt that Newton held the {\em accurate proportionality of the weight `to the masses of matter which they contain'} as inviolable}.

Whatever the case, Einstein promoted it to a truly {\em fundamental} cornerstone on which he erected his beautiful theoretical building: the Equivalence Principle (EP). Indeed, the postulated {\em exact} equality of the inertial and gravitational mass implies that, in a given constant and uniform gravitational field, all bodies move with the same acceleration in exactly the same way as they do in an uniformly accelerated reference frame removed from any external gravitational influence. In this sense, an uniformly accelerated frame in absence of gravity is  equivalent to an inertial frame in which a constant and uniform gravitational field is present.
It is important to stress that  the need of making the universality of the free fall, upon which the EP relies, compatible with the dictates of the SRT was not at all a trivial matter \cite{Renn} (cfr. Section \ref{inizio}), and the merit of keeping the law of free fall as a {\em fundamental} principle of a viable relativistic theory of gravitation  which could not reduce to a mere extension of the Newtonian theory to the SRT  must be fully ascribed to Einstein.  To better grasp the difficulties posed by such a delicate conceptual operation, let us think about an inertial reference frame $K$ in which two stones, differing by shape and composition, move under the action of a uniform gravitational field starting from the same height but with different initial velocities; for the sake of simplicity, let us assume  that, while one of the two stones is thrown horizontally with an initial velocity with respect to $K$, the other one falls vertically starting at rest \cite{Renn}. Due to the universality of free fall, both the stones reach the ground simultaneously. Let us, now, consider an inertial frame $K^{'}$ moving uniformly at a speed equal to the horizontal component of the velocity of the projectile; in this frame, the kinematics of the two objects gets interchanged: the projectile has no horizontal velocity so that now it falls vertically, while the stone at rest acquires an horizontal velocity making it move parabolically in the opposite direction with respect to $K^{'}$. According to the universality of the free fall, also in this case they should come to the rest at the same time. But this is in disagreement with the relativity of the simultaneity of the SRT. Moreover, another source of potential tension between the universality of the free fall and the SRT is as follows \cite{Renn}. According to the latter one, a change in the energy of a body corresponds to a change also in its inertial mass, which acts as a \virg{brake}. On the other hand, since the inertial mass is equivalent to the gravitational mass, which, instead, plays the role of \virg{accelerator}, the correct relativistic theory of gravitation necessarily implies that also the gravitational mass should depend in an exactly known way from the total energy of the body. Actually, other scientists like, e.g., Abraham \cite{Abr12} and Mie \cite{Mie13} were willing to discard the Galileo's law of universality of free fall to obtain a relativistic theory of gravitation.

The heuristic significance of the original form of the EP unfolded in the findings by Einstein that identical clocks ticks at different rates if placed at different points in a gravitational potential, a feature which was measured in a laboratory on the Earth's surface in 1960 \cite{moss} by means of the M\"{o}ssbauer effect which has recently received a general relativistic interpretation \cite{cordamos}, and the gravitational redshift of the spectral lines emitted at the Sun's surface with respect to those on the Earth, which was measured only in the sixties of the last century \cite{rs1} following the 1925 measurement with the spectral lines in the companion of Sirius \cite{rs2}. Furthermore, it  turned out  that the speed of light in  a gravitational field is variable, and thus light rays are deflected, as if not only an inertial mass but also a gravitational mass would correspond to any form of energy. Einstein \cite{Ein11} was also able to calculate the deflection of the apparent position of background stars due to the Sun's gravitational potential, although the value he found at that time was only half of the correct one later predicted with the final form of his GTR \cite{Ein15c} and measured in 1919 \cite{Dys20, Will15} (see Section \ref{fieldequations}). In 1912, he \cite{Ein12a, Ein12b} explored the possibility of gravitational lensing deriving the basic features of the lensing effect, which will be measured for the first time not until 1979 \cite{lente}. It must be noted \cite{Pau21, Renn} that this theory of the constant and uniform gravitational field went already {\em beyond} STR. Indeed, because of the dependence of the speed of light and the clock rates on the gravitational potential, STR definition of simultaneity and the Lorentz transformation themselves lost their significance (cfr. Section \ref{inizio}). {\em In this specific sense}, it can be said that STR can hold only in absence of a gravitational field.

The existence of non-uniformly accelerated reference frames like, e.g., those rotating with a time-dependent angular velocity ${\bds \Omega}(t)$, naturally posed the quest for a further generalization of the EP able to account for spatially and temporally varying gravitational fields as well.
The extension of the EP to arbitrarily accelerating frames necessarily implies, in principle, the existence of further, non-uniform, non-static (either stationary and non-stationary) and velocity-dependent gravitational effects, as guessed by Einstein \cite{EinGM, Ein13, EinBes13}. They were later fully calculated by Einstein himself \cite{Ein17} and others \cite{LT18, thir0, thir1, thir2, Mash93, indu1, indu2} with the final form of the GTR (see Section \ref{fieldequations} and \cite{Mash84, pfi1, pfi2} for critical analyses of the seminal works), which could not be encompassed by the gravito-static Newtonian framework. Indeed, it must be recalled that the inertial acceleration experienced by a body (slowly) moving with velocity ${\bds v}^{'}$ with respect to a rotating frame $K^{'}$ is
 \eqi {\bds a}^{'}_{\Omega} = 2{\bds\Omega}{\bds\times}{\bds v}^{'} + \dot{\bds\Omega}{\bds\times}{\bds r}^{'} + {\bds\Omega}{\bds\times}\ton{{\bds\Omega}{\bds\times}{\bds r}^{'}}\eqf At least to a certain extent, such new gravitational effects, some of which have been measured only a few years ago \cite{gpb, ciufo, iorioreview, renzreview}, might be considered as reminiscent of the Machian relational conceptions of mechanics \cite{Rin94, Bondi97, Rin97, Mach}.

 Such a generalization of the EP to arbitrary gravitational fields lead Einstein to reformulate it as follows: in any infinitesimal spacetime region (\emph{i.e}., sufficiently small to neglect either spatial and temporal variations of gravity throughout it), it is always possible to find a suitable non-rotating coordinate system $K_0$ in which any effect of gravity on either test particle motions and any other physical phenomena is absent. Such a local coordinate system can ideally be realized by a sufficiently small box moving in the gravitational field freely of any external force of non-gravitational nature. Obviously, it appeared natural to assume the validity of STR in $K_0$ in such a way that all the reference frames connected to it by a Lorentz transformation are physically equivalent.
{\em In this specific sense}, it could be said that the Lorentz covariance of all physical laws is still valid in the infinitely small.

At this point, still relying upon the EP, it remained to construct a theory valid also for arbitrarily varying gravitational fields by writing down the differential equations connecting the gravitational potential, assumed as state variable, with the matter-energy sources and requiring their covariance with respect to a fully general group of transformations of the spacetime coordinates.
\subsection{Predictions of the Equivalence Principle}\lb{eppredi}

A step forward was done in 1914 when, in collaboration with Grossmann, Einstein \cite{EinGros14}, on the basis of the Riemannian theory of curved manifolds, was able to introduce the ten coefficients $g_{\mu\nu}$ of the symmetric metric tensor $\tensorsym{g}$ by writing down the square of the spacetime line element $(ds)^2$ between two infinitely near events in arbitrary curvilinear coordinates $x^{\mu}$ as \eqi (ds)^2 = g_{\mu\nu} dx^{\mu}dx^{\nu}\eqf

As a consequence, the equations of motion of a test particle, the energy-momentum theorem and the equations of the electromagnetism {\em in vacuo} were simultaneously written in their generally covariant ultimate form. In particular, from the right-hand-side of the geodesic equation of motion of a test particle

\eqi \rp{d^2 x^{\alpha}}{ds^2} = -\Gamma^{\alpha}_{\beta\varrho}\rp{d x^{\beta}}{ds}\rp{dx^{\varrho}}{ds}\eqf where the Christoffel symbols \eqi \Gamma^{\alpha}_{\beta\varrho} \doteq \rp{1}{2}g^{\alpha\sigma}\ton{\rp{\partial g_{\sigma\beta}}{\partial x^{\varrho}} + \rp{\partial g_{\sigma\varrho}}{\partial x^{\beta}} - \rp{\partial g_{\beta\varrho}}{\partial x^{\sigma}}}\eqf are constructed with the first derivatives of $g_{\mu\nu}$, it was possible to straightforwardly identify the components of $\tensorsym{g}$ as the correct state variables playing the role of the Newtonian scalar potential $\Phi$.
Indeed, to a first-order level of approximation characterized by neglecting terms quadratic in $v/c$ and the squares of the deviations of the $g_{\mu\nu}$ from their STR values
\begin{align}
\eta_{00} &= +1\\ \nonumber
\eta_{ij} &= -\delta_{ij}
\end{align}
the geodesic equations of motion for the spatial coordinates become
\eqi \rp{d^2 x^i}{dt^2} = -c^2 \Gamma^i_{00}\eqf Furthermore, if the gravitational field is assumed static or quasi-static and the time derivatives can be neglected, the previous equations reduce to \eqi \rp{d^2 x^i}{dt^2} = \rp{c^2}{2}\rp{\partial g_{00}}{\partial x^i}\eqf By posing \eqi\Phi\doteq -\rp{1}{2}c^2\ton{g_{00} - 1}\eqf so that \eqi g_{00}= 1 - \rp{2\Phi}{c^2}\eqf the Newtonian acceleration is obtained. The additive constant up to which the potential is defined is fixed in such a way that $\Phi$ vanishes when $g_{00}$ assumes its STR value $\eta_{00}$. It is worthwhile remarking that, to the level of approximation adopted, only $g_{00}$ enters the equations of motion, although the deviations of the other metric coefficients from their STR values may be of the same order of magnitude. It is this circumstance that allows to describe, to a first order approximation, the gravitational field by means of a single scalar potential.

In analogy with the geodesic equations of motion for a test particle, also those for the propagation of electromagnetic waves followed. Indeed, the worldlines of light rays are, thus, geodesics curves of null~length
\begin{align}
(ds)^2 &= 0\\ \nonumber
\rp{d^2 x^{\alpha}}{d\lambda^2} &= -\Gamma^{\alpha}_{\beta\varrho}\rp{d x^{\beta}}{d\lambda}\rp{dx^{\varrho}}{d\lambda}
\end{align}
 where $\lambda$ is some affine parameter.

The components of the metric tensor $\tensorsym{g}$ are not assigned independently of the matter-energy distributions, being determined by field equations.

A further consequence of EP and the fact that, to the lowest order of approximation, $g_{00}$ is proportional to the Newtonian potential $\Phi$ is that, in general, it is possible to predict the influence of the  gravitational field on clocks even without knowing all the  coefficients $g_{\mu\nu}$; such an influence is actually determined by $g_{00}$ through \eqi d\tau = \sqrt{g_{00}} dt\eqf where $\tau$ is the reading of a clock at rest. Instead, it is possible to predict the behaviour of measuring rods only knowing all the other coefficients $g_{0i},~ g_{ik}$. Indeed, it turns out that the square of the distance $dl$ between two nearby points in the 3-dimensional space is given by \cite{landau} \eqi(dl)^2 = \ton{-g_{jh} + \rp{g_{0j}g_{0h}}{g_{00}}}dx^j dk^h\eqf Thus, the field  $\tensorsym{g}$ determines not only the gravitational field, but also the behaviour of clocks and measuring rods, \emph{i.e}., the chronogeometry of the 4-dimensional spacetime which contains the geometry of the ordinary 3-dimensional space as a particular case. Such a fusion of two fields until then completely separated-metric and gravitation-should be regarded as a major result of GTR, allowing, in principle, to determine the gravitational field just from local measurements of distances and time intervals.
\section{The Field Equations for the Metric Tensor and Their Physical Consequences}\lb{fieldequations}
\vspace{-12pt}
\subsection{The Field Equations}

The differential equations for the $\tensorsym{g}$ tensor itself
followed in 1915 \cite{Ein15a, Ein15b, Ein15d}.

The tortuous path \cite{Renn} which lead to them can be sketchily summarized as follows \cite{Pau21}.
According to the EP, the gravitational mass of a body is exactly equal to its inertial mass and, as such, it is proportional to the total energy content of the body. The same must, then, hold also in a given gravitational field for the force experienced by a body which is proportional to its (passive) gravitational mass. It is, thus, natural to assume that, conversely, only the energy possessed by a material system does matter, through its (active) gravitational mass, as for as its gravitational field is concerned. Nonetheless, in STR the energy density is not characterized by a scalar quantity, being, instead, the $00$ component of the so-called stress-energy tensor $\tensorsym{T}$.  It follows that also momentum and stresses intervene on the same footing as energy itself. These considerations lead to the assumption that no other material state variables than the components $T_{\mu\nu}$ of $\tensorsym{T}$  must enter the gravitational field equations.
Moreover, in analogy with the Poisson equation, $\tensorsym{T}$ must be proportional to a differential expression $\tensorsym{G}$ of the second order containing only the state variables of the gravitational field, \emph{i.e}., the components of the metric tensor $\tensorsym{g}$; because of the required general covariance, $\tensorsym{G}$ must be a tensor as well. The most general expression for it turned out to be \eqi G_{\mu\nu}= c_1 R_{\mu\nu} + c_2 g_{\mu\nu}R + c_3 g_{\mu\nu}\eqf where $\tensorsym{R}$ is the contracted curvature tensor whose components are \eqi R_{\mu\nu}= \rp{\partial\Gamma^{\alpha}_{\mu\alpha}} {\partial x^{\nu}} - \rp{\partial\Gamma^{\alpha}_{\mu\nu}}{\partial x^{\alpha}} + \Gamma^{\beta}_{\mu\alpha}\Gamma^{\alpha}_{\nu\beta} - \Gamma^{\alpha}_{\mu\nu}\Gamma^{\beta}_{\alpha\beta}\eqf and $R$ is its invariant trace.
The coefficients $c_1,~c_2,~c_3$ were determined by imposing that the stress-energy tensor satisfies the energy-momentum conservation theorem. By neglecting the third term in $\tensorsym{G}$, which usually plays a negligible role in the effects which will be discussed in this Section (see Section \ref{cosmo} for phenomena in which it may become relevant), the Einstein field equations became \cite{Ein15a, Ein15b}
\eqi\tensorsym{G} = -\varkappa \tensorsym{T}\eqf with \eqi G_{\mu\nu}= R_{\mu\nu}-\rp{1}{2}g_{\mu\nu} R\eqf and   $\varkappa$ is a constant which is determined by comparison with the Newtonian Poisson equation. By contraction, one gets \eqi R= \varkappa T\eqf where $T$ is the trace of $\tensorsym{T}$, so that
\eqi R_{\mu\nu} = -\varkappa\ton{T_{\mu\nu}-\rp{1}{2}g_{\mu\nu} T}\eqf This is the generally covariant form of the gravitational field equations to which, after many attempts, Einstein came in 1915 \cite{Ein15d}.

The same field equations were obtained elegantly by Hilbert through a variational principle \cite{Hil1}. On the reciprocal influences between Einstein and Hilbert in the process of obtaining the GTR field equations and an alleged priority dispute about their publication, see \cite{HilEin}.

It should be noted \cite{Pau21} that GTR, {\em per se}, yields neither the magnitude nor the sign (attraction or repulsion of the gravitational interaction) of $\varkappa$ which are, instead, retrieved from the observations.
For weak and quasi-static fields generated by pressureless, extremely slowly moving matter of density $\rho$, the right-hand-side of the field equation for the $00$ component becomes \eqi -\rp{1}{2}\varkappa c^2\rho\eqf indeed, the only non-vanishing component of the matter stress-energy tensor is \eqi T_{00}=\rho c^2\eqf so that \eqi T = -\rho c^2\eqf Since the time derivatives and the products of the Christoffel symbols can be neglected, the $00$ component of the Ricci tensor reduces to \eqi R_{00} = \rp{1}{2}{\bds\nabla}^2 g_{00} = -\rp{{\bds\nabla}^2\Phi}{c^2}\eqf Thus, it is \eqi{\bds\nabla}^2\Phi = \rp{1}{2}\varkappa c^4\rho\eqf the Poisson equation really holds. A comparison with the Newtonian equation tells that $\varkappa$ is positive, being equal to \eqi\varkappa = \rp{8\pi G}{c^4} = 2\times 10^{-43}~\textrm{kg}^{-1}~\textrm{m}^{-1}~\textrm{s}^2\eqf the spacetime can, thus, be assimilated to an extremely rigid elsatic medium.
\subsection{First Predictions of the Theory and Confrontation with Observations}

In the same year \cite{Ein15c}, Einstein readily employed his newborn theory to successfully explain the long-standing issue of the anomalous perihelion precession of Mercury \cite{lever}. To this aim, and also in order to derive the correct value of the deflection of a light ray grazing the Sun's limb \cite{Ein15c} through the Fermat principle, it was necessary to know not only the coefficient $g_{00}$ of the gravitational field of a point mass, as in the Newtonian approximation, but also the other metric coefficients $g_{ij}$. Since the spacetime outside a spherical body is isotropic, the off-diagonal metric coefficients $g_{0i}$ are identically zero: otherwise, they would induce observable effects capable of distinguishing between, e.g., two opposite spatial directions (see Section \ref{eppredi}). Moreover, it was also required to approximate $g_{00}$ itself to a higher order. Einstein \cite{Ein15c} solved that problem by successive approximations. The {\em exact} vacuum solution was obtained one year later by Schwarzschild \cite{schw} and, independently, Droste \cite{droste}; their results are virtually indistinguishable from those of Einstein. Relevant simplifications were introduced one year later by Weyl \cite{weyl}, who used cartesian coordinates instead of spherical ones, and worked on the basis of the action principle instead of recurring to the differential equations for the field $\tensorsym{g}$.
Schwarzschild~\cite{dentro1} extended the validity of his solution also to the interior of a material body modelled as a sphere of incompressible fluid.
Having in hand this exact solution of the Einstein field equations revolutionized the successive development of GTR. Indeed, instead of
dealing only with small weak-field corrections to Newtonian gravity, as Einstein had initially imagined would be the case, fully nonlinear features of the theory such as gravitational collapse and singularity formation could be studied, as it became clear decades later.
 About the Schwarzschild solution, the Birkhoff's Theorem \cite{birk} was proved in 1923. According to it, even without the assumption of staticity, the Schwarzschild metric is the {\em unique} vacuum solution endowed with spherically symmetry. As a consequence, the external field of a spherical body radially pulsating or radially imploding/exploding is not influenced at all by such modifications of its~source.

The successful explanation of the anomalous perihelion precession of Mercury was a landmark for the validity of GTR since, as remarked in \cite{Brush, Weinb}, it was a successful {\em retrodiction} of an effect which was known for decades. In particular, Weinberg wrote \cite{Weinb}: ``It is widely supposed that the true test of a theory is in the comparison of its predictions with the results of experiment. Yet, with the benefit of hindsight, one can say today that Einstein's successful explanation in 1915 of the previously measured anomaly in Mercury's orbit was a far more solid test of general relativity than the verification of his calculation of the deflection of light by the sun in observations of the eclipse of 1919 or in later eclipses. That is, in the case of general relativity a {\em retrodiction}, the calculation of the already-known anomalous motion of Mercury, in fact provided a more reliable test of the theory than a true {\em prediction} of a new effect, the deflection of light by gravitational fields.

I think that people emphasize prediction in validating scientific theories because the classic attitude  of commentators on science is not to trust the theorist. The fear is that the theorist adjusts his or her theory to fit whatever experimental facts are already known, so that for the theory to fit to these facts is not a reliable test of the theory.

But [\ldots] no one who knows anything about how general relativity was developed by Einstein, who at all follows Einstein's logic, could possibly think that Einstein developed general relativity in order to explain this precession. [\ldots] Often it is a successful {\em prediction} that one should really distrust. In the case of a true prediction, like Einstein's prediction of the bending of light by the sun, it is true that the theorist does not know the experimental result when she develops the theory, but on the other hand the experimentalist does know about the theoretical result when he does the experiment. And that can lead, and historically has led, to as many wrong turns as overreliance on successful retrodictions. I repeat: it is not that experimentalists  falsify their data. [\ldots] But experimentalists who know the result that they are theoretically supposed to get naturally find it difficult to stop looking for observational errors when they do not get that result or to go on looking for errors when they do. It is a testimonial to the strength of character of experimentalists that they do not always get the results they expect''.

The final work of Einstein on the foundations of GTR appeared in 1916 \cite{Ein16}.

In the same year, de Sitter \cite{des} was able to derive a further consequence of the static, spherically symmetric spacetime of the Schwarzschild solution: the precession of the orbital angular momentum of a binary system, thought as a giant gyroscope, orbiting a non-rotating, spherical body such as in the case of the Earth-Moon system in the Sun's field. Some years later, Schouten \cite{sciou} and Fokker \cite{fok} independently obtained the same effect by extending it also to spin angular momenta of rotating bodies. Such an effect is mainly known as de Sitter or geodetic precession. It was measured decades later in the field of the Sun by accurately tracking the orbit of the Earth-Moon system with the Lunar Laser Ranging technique \cite{llr1, llr2}, and in the field of the Earth itself with the dedicated Gravity Probe B (GP-B) space-based experiment \cite{gpb} and its spaceborne gyroscopes.

In 1964 \cite{shap}, Shapiro calculated a further prediction of the static Schwarzschild spacetime: The temporal delay, which since then bears his name, experienced by travelling electromagnetic waves which graze the limb of a massive body as the Sun in a back-and-forth path to and from a terrestrial station after having been sent back by a natural or artificial body at the superior conjunction with our planet. In its first successful test performed with radar signals \cite{shaptest}, Mercury and Venus were used as reflectors. Latest accurate results \cite{bertotti} relied upon the Cassini spacecraft en route to Saturn.

\subsection{The General Approximate Solution by Einstein}

In 1916 \cite{Ein16b}, Einstein, with a suitable approximation method,  was able to derive the field generated by bodies moving with arbitrary speeds, provided that their masses are small enough. In this case, the $g_{\mu\nu}$ differ slightly from the STR values  $\eta_{\mu\nu}$, so that the squares of their deviations $h_{\mu\nu}$ with respect to the latter ones can be neglected, and it is possible to keep just the linear part of the field equations. Starting from their form \cite{Ein15a, Ein15b} \eqi R_{\mu\nu} - \rp{1}{2}g_{\mu\nu}R = - \varkappa T_{\mu\nu}\eqf  working in the desired approximation, they can be cast into a linearized form in terms of the auxiliary state variables \eqi{\overline{h}}_{\mu\nu} \doteq h_{\mu\nu} - \rp{1}{2}\delta_{\mu}^{\nu} h\eqf
 where $\delta_{\mu}^{\nu}$ is the Kronecker delta, and $h$ is the trace of $\tensorsym{h}$ which is a tensor only with respect to the Lorentz transformations. A further simplification can be obtained if suitable spacetime coordinates, satisfying the gauge condition

 \eqi\rp{\partial {\overline{h}}_{\alpha\beta}}{\partial x^{\beta}} = 0\eqf known as Lorentz gauge (or Einstein gauge or Hilbert gauge or de Donder gauge or Fock gauge), are adopted.
The resulting differential equations for the  state variables ${\overline{h}}_{\mu\nu}$ are \eqi\square {\overline{h}}_{\mu\nu} = - 2\varkappa T_{\mu\nu}\eqf which is the inhomogeneous wave equation; $\square$ is the STR form of the d'Alembertian operator. The usual method of the retarded potentials allows to obtain \eqi{\overline{h}}_{\mu\nu} = \rp{\varkappa}{2\pi}\int \rp{T_{\mu\nu}\ton{x^{'},y^{'},z^{'},t - r/c}}{r}dx^{'}dy^{'}dz^{'}\eqf Among other things, it implies that the action of gravity propagates to the speed of light: a quite important results which, some years ago, was the subject of dispute \cite{Will03, Carl05, Kop05} boosted by the interpretation of certain VLBI measurements of the time delay suffered at the limb of Jupiter by electromagnetic waves from distant quasars \cite{Kop, Kop2}.

\subsubsection{Gravitational Waves}

The form of the gravitational waves in empty regions follows from the Lorentz gauge condition  and the inhomogeneous wave equation by posing $\tensorsym{T} = 0$: it was studied by Einstein in \cite{Ein18}, where he also calculated the emission and the absorption of gravitational waves. It turned out that, when  oscillations or other movements take place in a material system, it emits gravitational radiation in such a way that the total power emitted along all spatial directions is determined by the third temporal derivatives of the system's moment of inertia \eqi I_{ij}=\int \rho x^i x^j dx^1 dx^2 dx^3\eqf Instead, when a gravitational wave impinges on a material system whose size is smaller than the wave's wavelength, the total power absorbed is determined by the second temporal derivatives of its moment of inertia \cite{Ein18}.

Gravitational waves were {\em indirectly} revealed for the first time \cite{gw1, gw3, gw2} in the celebrated Hulse-Taylor binary pulsar PSR B1913+16 \cite{HT, DamPSR}. {\em Direct} detection (some of) their predicted effects in both terrestrial \cite{groundgw1, groundgw2, groundgw3, groundgw4, groundgw5, groundgw6} and space-based laboratories \cite{spacegw1, spacegw2, spacegw3, spacegw4, spacegw5, spacegw6, spacegw7} from a variety of different astronomical and astrophysical sources \cite{source1}, relentlessly chased by at least fifty years since the first proposals by Gertsenshtein and Pustovoit \cite{russigw} of using interferometers and the pioneering attempts by J. Weber~\cite{weber1} with its resonant bars \cite{weber0}, is one of the major challenges of the current research in relativistic physics~\cite{damgw, cordagw}.

Conversely, by assuming their existence, they could be used, in principle, to determine key parameters of several extreme astrophysical and cosmological scenarios which, otherwise, would remain unaccessible to us because of lack of electromagnetic waves from them \cite{satyap} by establishing an entirely new \virg{Gravitational Wave Astronomy} \cite{damgw, gwastro}.
A recent example \cite{Gasp0} is given by the possibility that the existence of primordial gravitational waves may affect the polarization of the electromagnetic radiation  which constitutes the so-called Cosmic Microwave Background (CMB), discovered in 1965~\cite{cmb}. In this case, the polarizing effect of gravity is indirect since the field of the gravitational waves does not directly impact the polarization of CMB, affecting, instead, the anisotropy of the spatial distribution of CMB itself. Indeed, the polarization of CMB is a direct consequence of the scattering of the photons of the radiation with the electrons and positrons which formed the primordial plasma, existing in the primordial Universe at the so-called decoupling era \cite{Spergel}. At later epochs, when the temperature fell below 3000 K$^{\circ}$, the radiation decoupled from matter, photons and electrons started to interact negligibly, and the polarization got \virg{frozen} to the values reached at the instant of decoupling.
Thus, mapping the current CMB's polarization state  has the potential of providing us with direct information of the primordial Universe, not contaminated by the dynamics of successive evolutionary stages. In particular, it turns out that the presence of metric fluctuations of tensorial type, \emph{i.e}., of gravitational waves, at the epoch in which the CMB radiation interacted with the electrons of the cosmic matter  getting polarized, may have left traces in terms of polarization modes of B type \cite{B1, B2}. They could be currently measurable, provided that the intensity of the cosmic background of gravitational waves is strong enough. An example of cosmic gravitational background able to produce, in principle, such an effect is represented by the relic gravitational radiation produced during the inflationary epochs. The gravitational waves produced in this way are distributed over a very wide frequency band $\Delta\omega(t)$ which is generally time-dependent. In order to characterize the intensity of such relic gravitational waves, it turns out convenient to adopt the spectral energy density \eqi\varepsilon_h(\omega,~t)\doteq \rp{d\varepsilon(t)}{d\ln\omega}\eqf defined as the energy density $\varepsilon(t)$ per logarithmic interval of frequency, normalized to the critical energy density $\varepsilon_{\rm crit}$ (see Section \ref{flrw}), \emph{i.e}., the dimensionless variable \eqi\upOmega_h(\omega,~t)\doteq \rp{1}{\varepsilon_{\rm crit}}\rp{d \varepsilon}{d\ln\omega}\eqf The simplest inflationary models yield power-law signatures for it. In 2014 \cite{flop1}, the BICEP2  experiment at the South Pole seemed to have successfully revealed the existence of the B modes; the measured values seemed approximately in agreement-at least in the frequency band explored by BICEP2-with a cosmic gravitational radiation background corresponding to the aforementioned power-law models.
More recently \cite{flop2}, a joint analysis of data from ESA's Planck satellite and the ground-based BICEP2 and Keck Array experiments did not confirm such a finding.
\subsubsection{The Effect of Rotating Masses}

The previously mentioned solution ${\overline{h}}_{\mu\nu}$ of the inhomogenous wave equation in terms of the retarded potentials  was used by Thirring \cite{thir1, thir2, Mash84} to investigate, to a certain extent, the relative nature of the centrifugal and Coriolis fictitious forces arising in a rotating coordinate system with respect to another one connected with the static background of the fixed stars. Indeed, according to a fully relativistic point of view, they should also be viewed as gravitational effects caused by the rotation of the distant stars with respect to a fixed coordinate system. At first sight, it may seem that such a possibility is already included in the theory itself in view of the covariance of the field equations. Actually, it is not so because the boundary conditions at infinite distance play an essential role in selecting, {\em de facto}, some privileged coordinate systems, in spite of a truly \virg{relativistic} spirit with which the theory should be informed. In other words, although the equations of the theory are covariant, the choice of the boundary conditions at spatial infinity, which are distinct from and independent of the field equations themselves, would pick up certain coordinate systems with respect to others, which is a conceptual weakness of an alleged \virg{generally relativistic} theory. Thus, Thirring \cite{thir1} did not aim to check the full equivalence of the gravitational effects of the rotation of the whole of the distant stars of the Universe with those due to the rotation of the coordinate system with respect to them, assumed fixed. Indeed, he considered just a rotating hollow shell of finite radius $D$ and mass $M$, so to circumvent the issue of the boundary conditions at infinite distance by setting the spacetime metric tensor there equal to the Minkowskian one. By assuming $M$ small with respect to the whole of the fixed stars, so to consider the departures of the $g_{\mu\nu}$ coefficients  from their STR values $\eta_{\mu\nu}$ small inside the shell, the application of the previously obtained Einsteinian expression for ${\overline{h}}_{\mu\nu}$ to the shell yielded that a test particle inside the hollow space inside it is affected by accelerations which are formally identical to the centrifugal and Coriolis ones, apart from a multiplicative scaling dimensionless factor as little as \eqi \rp{GM}{c^2 D}\eqf

This explains the failures by Newton \cite{scoglio} in attributing the centrifugal curvature of the free surface of water in his swirling bucket to the relative rotation of the bucket itself and the water, and the Friedl\"{a}nder brothers \cite{twins} who unsuccessfully attempted to detect centrifugal forces inside a heavy rotating flywheel.

Another application of the approximate solution ${\overline{h}}_{\mu\nu}$ of the inhomogeneous wave equation allowed to discover that, while in either GTR and the Newtonian theory the gravitational field of a static, spherical body is identical to that of a point mass \cite{birk}, it is not so-in GTR-if the body rotates. Indeed, Einstein~\cite{Ein17}, Thirring and Lense \cite{LT18} calculated the (tiny) precessions affecting the orbits of test particles as natural satellites and planets moving in the field of rotating astronomical bodies such as the Sun and some of its planets. Such a peculiarity of the motion about mass-energy currents, universally known as \virg{Lense-Thirring effect} by historical tradition (cfr. \cite{pfi1} for a critical historical analysis of its genesis), was subjected to deep experimental scrutiny in the last decades \cite{ciufo, iorioreview, renzreview}.

In the sixties of the twentieth century, another consequence of the rotation of an astronomical body was calculated within GTR: the precession of an orbiting gyroscope \cite{pugh, schiff}, sometimes dubbed as \virg{Pugh-Schiff effect}. The GP-B experiment \cite{gpb0}, aimed to directly measure also such an effect in the field of the Earth,  was successfully completed a few years ago \cite{gpb}{, although the final accuracy obtained ($\sim$19\%) was worse than that expected ($\sim$1\% or better) }.
\subsection{Black Holes and Other Physically Relevant Exact Solutions of the Field Equations}
\vspace{-12pt}
\subsubsection{The Reissner-Nordstr\"{o}m Metric}
In 1916, Reissner \cite{reiss} solved the coupled Einstein-Maxwell field equations and found the metric which describes
the geometry of the spacetime surrounding a  pointlike electric charge $Q$. One year later, Weyl \cite{weyl} obtained the same metric from a variational action principle. In 1918, Nordstr\"{o}m \cite{nords}, generalized it to the case of a spherically symmetric charged body. The metric for a non-rotating charge distribution is nowadays known as the Reissner-Nordstr\"{o}m metric; in the limit $Q\rightarrow 0$, it reduces to the Schwarzschild solution.

The physical relevance of the Reissner-Nordstr\"{o}m metric in astronomical and astrophysical scenarios depends on the existence of macroscopic bodies stably endowed with net electric charges.
\subsubsection{Black Holes}

One of the consequences of the vacuum Schwarzschild solution was that it predicts the existence of a surface of infinite red-shift at \eqi r=r_{\rm g}\doteq \rp{2 GM}{c^2}\eqf  Thus, if, for some reasons, a body could shrink so much to reduce to such a size, it would disappear from the direct view of distant observers, who would not be anymore able to receive any electromagnetic radiation from such a surface, later interpreted as a spatial section of an \virg{event horizon} \cite{rind, finkel, brill}. A \virg{frozen star}, a name common among Soviet scientists from 1958 to 1968 \cite{frozen1, frozen2}, would have, then, formed, at least from the point of view of an external observer. In 1968 \cite{frozen1, buco}, Wheeler renamed such objects with their nowadays familiar appellative of \virg{black holes} \cite{ruff}.

In fact, both Eddington in 1926 \cite{edd26} and Einstein in 1939 \cite{Ein39}, although with arguments at different levels of soundness, were firmly convinced that such bizarre objects could not form in the real world. Instead, in 1939 \cite{oppen1}, Oppenheimer and Snyder demonstrated that, when all the thermonuclear sources of energy are exhausted, a sufficiently heavy star will unstoppably collapse beyond its Schwarzschild radius to end in a spacetime singularity. The latter one is not to be confused with the so-called \virg{Schwarschild singularity} occurring in the Schwarzschild metric at $r=r_{\rm g}$, which was proven in 1924~\cite{art1} to be unphysical, being a mere coordinate artifact; nonetheless, it took until 1933 for Lema\^{\i}tre to realize it \cite{tooft}.
In 1965, Penrose \cite{penro1}, in his first black hole singularity theorem, demonstrated that the formation of a singularity at the end of a gravitational collapse was an inevitable result, and not just some special feature of spherical symmetry.
A black hole is the  4-dimensional spacetime region  which represents the future of an imploding star: it insists on the 2-dimensional spatial critical surface determined by the star's Schwarzschild radius.  The 3-dimensional spacetime hypersurface delimiting the black hole, \emph{i.e}., its event horizon, is located in correspondence of the critical surface \cite{brill}.
\subsubsection{The Kerr Metric}

In 1963, the third physically relevant {\em exact} vacuum solution of the Einstein field equations was found by Kerr \cite{kerr}. It describes the spacetime metric outside a rotating source endowed with mass $M$ and proper angular momentum $J$. It was later put in a very convenient form by Boyer and Lindquist \cite{boli}.
At that time, it was generally accepted that a spherical star would collapse to a black hole described by the Schwarzschild metric. Nonetheless, people was wondering if such a dramatic fate of a star undergoing gravitational collapse was merely an artifact of the assumed perfect spherical symmetry. Perhaps, the slightest angular momentum would halt the collapse before the formation of an event horizon, or at least before the
formation of a singularity. In this respect, finding a metric for a rotating star would have been quite valuable.

Contrary to the Schwarzschild solution \cite{dentro1}, the Kerr one has not yet been satisfactorily extended to the interior of any realistic matter-energy distribution, despite several attempts over the years \cite{nodentro}. Notably, according to some researchers \cite{fuffa0, fuffa1, bigkerr}, this limit may have no real physical consequences since the exterior spacetime of a rotating physically likely source is {\em not} described by the Kerr metric whose higher multipoles, according to the so-called \virg{no-hair} conjecture \cite{hair0, hair1}, can all be expressed in terms of $M$ and $J$ \cite{multi1, multi2}, which is not the case for a generic rotating star \cite{multiveri}. Moreover, the Kerr solution does not represent the  metric during any realistic gravitational collapse; rather, it yields the asymptotic metric at late times as whatever dynamical process produced the black hole settles down, contrary to the case of a non-rotating collapsing star whose exterior metric is described by the Schwarzschild metric at all times. The Birkhoff's Theorem \cite{birk} does not hold for the Kerr metric.

The enormous impact that the discovery by Kerr has had in the subsequent fifty years on every subfield of GTR and astrophysics as well is examined in \cite{bigkerr}; just as an example, it should be recalled that, at the time of the Kerr's discovery, the gravitational collapse to a Schwarzschild black hole had difficulty in explaining the impressive energy output of quasars, discovered and characterized just in those \mbox{years~\cite{qso1, qso2}}, because of the \virg{frozen star} behavior for distant observers. Instead, the properties of the event horizon were different with rotation taken into account. A comparison of the peculiar features of the Schwarzschild and the Kerr solutions can be found in \cite{compa}.
\subsubsection{The Kerr-Newman Metric}

In 1965 \cite{kerrnewman0}, a new {\em exact} vacuum solution of the Einstein-Maxwell equations of GTR appeared: the Kerr-Newman metric \cite{kerrnewman1}. It was obtained from the Reissner-Nordstr\"{o}m metric by a complex transformation algorithm \cite{algo} without integrating the field equations, and is both the spinning generalization of Reissner-Nordstr\"{o}m and the electrically charged version of the Kerr metric.
 Such solutions point towards the possibility that charged and rotating bodies can undergo gravitational collapse to form black holes just as in the uncharged, static case of the Schwarzschild metric.

Leaving the issue of its physical relevance for astrophysics applications out of consideration, the Kerr-Newman metric is the most general static/stationary black hole solution to the Einstein-Maxwell equations. Thus, it is of great importance for theoretical considerations within the mathematical framework of GTR and beyond. Furthermore, understanding this solution also provides valuable insights into the other black hole solutions, in particular the Kerr metric.

\section{Application to Cosmology}\lb{cosmo}
\vspace{-12pt}
\subsection{Difficulties of Newtonian Cosmologies}

The birth of modern cosmology might be dated back to the correspondence between Newton and Bentley in the last decade of the seventieth century \cite{Ry}, when the issue of the applicability of Newtonian gravitational theory to a static, spatially infinite (Euclidean) Universe uniformly filled with matter was tackled. In four letters to R. Bentley, Newton explored the possibility that matter might be spread  uniformly throughout an infinite space. To the Bentley' s suggestion that such an even distribution might be stable, Newton replied that, actually, matter would tend to collapse into large massive bodies. However, he apparently also thought that they could be stably spread throughout all the space. In particular, in his  letter of 10 December 1692, Newton wrote \cite{lettera10dic}: \virg{it seems to me that if [\ldots] all the matter in the Vniverse was eavenly scattered throughout all the heavens, $\&$ every particle had an innate gravity towards all the rest $\&$ the whole space throughout which this matter was scattered, was but finite: the matter on the outside of this space would by its gravity tend towards all the matter on the inside $\&$ by consequence fall down to the middle of the whole space $\&$ there compose one great spherical mass But if the matter was eavenly diffused through an infinite space, it would never convene into one mass but some of it convene into one mass $\&$ some into another so as to make an infinite number of great masses scattered at great distances from one another throughout all that infinite space.}

Connected with the possibility that matter would fill uniformly an infinite space, and, thus, indirectly with the application of Newtonian gravitation to cosmology, there was also the so-called Olbers paradox~\cite{olbers}, some aspects of which had been previously studied also by Kepler \cite{kepler}, \mbox{Halley \cite{halleya, halleyb}} and de Ch\'{e}seaux \cite{deces}. According to it, although the light from stars diminishes as the square of the distance to the star, the number of stars in spherical shells increases as the square of the shell's radius. As a result, the accumulated effect of the light intensity should make the night sky as bright as the surface of the Sun. In passing, the Olbers paradox touched also other topics which will become crucial in contemporary cosmology like the temporal infinity of the Universe and its material content, and its spatial infinity as well.
At the end of the nineteenth century, Seeliger \cite{see1} showed that, in the framework of the standard Newtonian theory, matter cannot be distributed uniformly throughout an infinite Universe. Instead, its density should go to zero at spatial infinity  faster than $r^{-2}$; otherwise, the force exerted on a point mass by all the other bodies of the Universe would be undeterminate because it would be given by a non-convergent, oscillating series. Later, Einstein \cite{EinCosmo} critically remarked that, if the potential was finite at large distances as envisaged by Seeliger \cite{see1} to save the Newtonian law, statistical considerations would imply a depopulation of the fixed stars ensemble, assumed initially in statistical equilibrium. The possibility of an infinite potential at large distances, corresponding to a finite or vanishing not sufficiently fast matter density, already ruled out by Seeliger himself, was excluded also by Einstein \cite{EinCosmo} because it would yield unrealistically fast speeds of the distant stars. Seeliger \cite{see2} demonstrated also that matter density could be different from zero at arbitrary distances if the standard Poisson equation was modified as \eqi{\bds\nabla}^2\Phi - \Lambda \Phi= 4\pi G\rho\eqf  It admits \eqi\Phi = -\rp{4\pi G\rho}{\Lambda}\eqf as a viable solution for a uniform matter density, thus making an evenly filled Universe stable.
For a discussion of the problems encountered by the Newtonian theory of gravitation to cosmology, see, e.g.,~\cite{norton}.

The inadequacy of Newtonian gravitation to cosmological problems can be also inferred in view of the modern discoveries concerning the expansion of the Universe over the eons (see Section \ref{cosmomodels}) which, in conjunction with the finite value of $c$, yielded to the notion of observable Universe.
As previously recalled in Section \ref{eppredi}, the gravitational interaction among macroscopic bodies can be adequately described, to the first approximation,  by the non-relativistic Newtonian model. Such an approximation is applicable over spatial scales ranging from laboratory to planetary, stellar, and galactic systems.
On the other hand \cite{Gasp}, the Newtonian model cannot be applied, not even to the first approximation, to correctly describe gravity over cosmological distances of the order the Hubble distance \eqi D_{\rm H}\doteq \rp{c}{H_0}\sim 10^{26}~\textrm{m}\eqf where \cite{planckS} \eqi H_0 = (67.3\pm 1.2)~ \textrm{km}~\textrm{s}^{-1}~\textrm{Mpc}^{-1}\eqf is the current value of the Hubble parameter (see Section \ref{flrw}), which fixes the maximum spatial distance accessible to current observations (the radius of the observable universe is proportional to $D_{\rm H}$ through a numerical coefficient which, according to the present-day cosmological parameters, is equal to $3.53$).
Indeed, the absolute value of the potential of the mass equivalent to the energy density $\varepsilon$ enclosed in a spherical volume of radius $\sim D_{\rm H}$ is \eqi\left|\Phi_{\rm H}\right| = \rp{4}{3}\rp{\pi G\varepsilon}{H_0^2}\eqf The condition of validity of the Newtonian approximation is that, for any test particle of mass $m$, the gravitational potential energy $m\left|\Phi_{\rm H}\right|$ resulting from the interaction with the cosmological mass of the observable Universe is much smaller than its rest energy $mc^2$. Instead, it turns out \cite{Gasp} \eqi\rp{4}{3}\rp{\pi G\varepsilon}{H_0^2 c^2}\sim 1\eqf  It follows that the Newtonian approximation  is not valid at the Hubble scale, and a correct dynamical description of the Universe  to cosmological scales must necessarily rely upon a relativistic theory of~gravity.
\subsection{Relativistic Cosmological Models}\lb{cosmomodels}

GTR, applied to cosmology for the first time in 1917 by Einstein himself \cite{EinCosmo}, was able to put such a fundamental branch of our knowledge on the firm grounds of empirical science.

In the following, we will try to follow the following terminological stipulations \cite{chiffon}. We will generally use the word \virg{Universe} to denote a model of the cosmological spacetime along with its overall matter-energy content; as we will see, the relativistic Universe is the space woven by time and weighed by all forms of energy (matter-either baryonic and non-baryonic-, radiation, cosmological constant). As such, the Universe has neither center nor borders, neither inside nor outside. Instead, by means of \virg{universe} we will denote the observable portion of the cosmological spacetime delimited by a cosmological horizon unavoidably set by the fact that all the physical means (electromagnetic and gravitational radiation, neutrinos, cosmic rays) by which we collect information from objects around us travel at finite speeds. Its spatial section is a centered on the Earth-based observer with a radius equal to \eqi 3.53 D_{\rm H}= 51.3~{\rm Gly} = 15.7~{\rm Gpc}\eqf
\subsubsection{The Static Einstein Model}

In 1917, Einstein \cite{EinCosmo}  showed that, following his field equations in their original form, it would not be possible to choose the boundary conditions in such a way to overcome simultaneously {the} depopulation and the observed small stellar velocities issues. Instead, in principle, it is mathematically possible to modify them in as much as the same way as it was doable with the Poisson equation by introducing a $\Lambda$ term which yielded \eqi R_{\mu\nu} +\Lambda g_{\mu\nu}= -\varkappa\ton{T_{\mu\nu}-\rp{1}{2}g_{\mu\nu} T}\eqf Some years later, Cartan \cite{cartan22} demonstrated that the most general form of the Einstein field equations necessarily implies the $\Lambda$ term. It turned out that a Universe uniformly filled with \textcolor{black}{constant} matter density $\rho$ and non-vanishing \eqi \Lambda_{\textrm E} = \rp{4\pi G\rho}{c^2}\eqf would rest in equilibrium. Moreover, since it would be spatially closed with
\begin{align}
g_{00} &= 1\nonumber \\
g_{0i} &=0\\
g_{ij} &= -\qua{\delta_{ij} + \rp{x_i x_j}{S^2 - \ton{x_1^2 + x_2^2 + x_3^2} }}\nonumber
\end{align}
and radius $S$ connected with $\Lambda$ by \eqi\Lambda =  \rp{1}{S^2}\eqf there would not be the need of choosing suitable boundary conditions at infinity, thus removing the aforementioned \virg{non-relativistic} drawback of the theory (see Section \ref{fieldequations}). It should be noted that if such a $4-$dimensional cylindrical Universe did not contain matter, there would not be any gravitational field,~\emph{i.e}., \eqi T_{\mu\nu}=0\eqf would imply \eqi g_{\mu\nu}=0\eqf

Thus, the postulate of the complete relativity of inertia would be met. In the Einstein spatially hyperspherical model, the spacetime trajectories of moving bodies and light rays wind around spirals on the surface of a cylinder in such a way that if one watched a spaceship moving away from her/him, it first would diminish in size but then would come back beginning to magnify again.  Thirteen years later, Eddington \cite{edd} showed that the static Einsteinian model is, actually, unstable.

It may be interesting to note \cite{torr} how the Einstein's Universe is, in fact, no less liable to the Olbers paradox than the Newtonian one; indeed, the light emitted by a star would endlessly circumnavigate the static spherical space until obstructed by another star.
\subsubsection{The de Sitter Model}

In 1917, de Sitter \cite{unides1, unides2} found a solution for the modified Einstein field equations with $\Lambda \neq 0$ yielding a 4-dimensional hyperbolic Universe
\begin{align} g_{\mu\nu} &= \rp{\eta_{\mu\nu}}{\ton{1- \rp{\Lambda}{12}\eta_{\alpha\beta}x^{\alpha} x^{\beta}}^2}\\ \nonumber
\Lambda &= \rp{3}{S^2}
\end{align} with non-zero gravitational field even in absence of matter, thus differing from the Einstein model. It allowed also a sort of spatial (and not material) origin of inertia, which would be relative to void space: a hypothetical single test particle existing in the otherwise empty de Sitter Universe would have inertia just because of $\Lambda$.

At the time of the Einstein and de Sitter models, there were not yet {\em compelling} means to observationally discriminate between them \cite{unides3}, although their physical consequences were remarkably different. Suffice it to say that the spacetime geometry of the de Sitter Universe implied that, although static, test particles  would have escaped far away because of the presence of the $\Lambda$ term, unless they were located at the origin. Such a recessional behaviour was known as \virg{de Sitter effect}.
As said by Eddington \cite{eddbook}, \virg{the de Sitter Universe contains motion without matter, while the Einstein Universe contains matter without motion}.

After having lost appeal with the advent of the genuine non-stationary Fridman-Lema\^{\i}tre  solutions (see Section \ref{fridy}), the de Sitter model was somewhat revamped in the framework of the inflationary phase characterized by an ultrafast expansion that it is believed to have occurred in the early stages of the universe \cite{staro, guth, linde}.
\subsubsection{The Fridman-Lema\^{\i}tre-Robertson-Walker Expanding Models}\lb{fridy}

In the twenties of the last century, the first truly non-static theoretical models of the Universe were proposed by Fridman \cite{fried1, fried2}. Indeed, he found new solutions of the Einstein field equations with $\Lambda$ representing spatially homogeneous and isotropic cosmological spacetimes filled with matter-energy modeled as a perfect fluid generally characterized only by time-varying density $\rho(t)$, and endowed with an explicitly time-dependent universal scaling factor $S(t)$ for the spatial metric having constant curvature $k=0,\pm1$ throughout all space. If $k = +1$, the 3-dimensional space is {\em spherical} and necessarily {\em finite} (as the hypersphere); if $k = 0$, it is {\em Euclidean}; if $k = -1$, it is {\em hyperbolic}. Euclidean and hyperbolic spaces can be either finite or infinite, depending on their topology \cite{topo1, topo2} which, actually, is {\em not} determined by the Einstein field equations governing only the dynamical evolution of $\rho(t),~S(t)$. Importantly, viable solutions exist also in absence of the cosmological $\Lambda$ term for all the three admissible values of the spatial curvature parameter $k$. The Einstein and de Sitter models turned out \cite{fried1} to be merely limiting cases of an infinite family of solutions of the Einstein field equations for a positive, time-varying matter density, any one of which would imply, at least for a certain time span, a general recession-or oncoming, since the solutions are symmetric with respect to time reversal-of test particle. According to their dynamical behaviour, the Fridman's models are classified as {\em closed} if they recollapse, {\em critical} if they expand at an asymptotically zero rate, and {\em open} if they expand indefinitely. In this respect, a spherical universe can be open if  $\Lambda$ is positive and large enough, but it cannot be infinite. Conversely, Euclidean or hyperbolic universes, generally open, can be closed if $\Lambda<0$; their finiteness or infiniteness depends on their topology, not on their material content. Fridman's simplifying assumptions were much weaker than those of either Einstein and de Sitter, so that they defined a much likelier idealization of the real world~\cite{torr}, as it turned out years later: indeed, the russian scientist was interested only in the mathematical aspects of the cosmological solutions of the Einstein equations.

Approximately in the same years, a body of observational evidence pointing towards mutual recessions of an increasingly growing number of extragalactic nebulae was steadily \mbox{accumulating~\cite{sliph1, sliph2, sliph3, hub, hum}} from accurate red-shifts measurements, probably unbeknownst to Fridman.  In 1929, Hubble \cite{hub} made his momentous claim that the line-of-sight speeds of the receding galaxies are proportional to their distances from the Earth. If, at first, the de Sitter model, notwithstanding its material emptiness, was regarded with more favor than the Einstein one as a possible explanation of the observed red-shifts of distant nebul{\ae}, despite the cautiousness by de Sitter himself \cite{unides2}, it would have been certainly superseded by the more realistic Fridman ones, if only they  had been widely known at that time (Fridman died in 1925). It may be that a role was played in that by the negative remark by Einstein about a claimed incompatibility of the non-stationary Fridman's models  with his field equations \cite{nein1}, later retracted by the father of GTR because of an own mathematical error in his criticism \cite{nein2}.

At any rate, in 1927, Lema\^{\i}tre \cite{lema}, who apparently did never hear of the Fridman's solutions, rederived them and applied them to the physical universe with the explicit aim of founding a viable explanation of the observed recessions of galaxies (the red-shifted nebul{\ae} had been recognized as extra-galactic objects analogous to our own galaxy in 1925 by Hubble \cite{hubgal}). Lema\^{\i}tre \cite{lema} also showed that the static solution by Einstein is unstable with respect to a temporal variation of matter density. Enlightened by the Hubble's discovery \cite{hub}, and, perhaps, also struck by the criticisms by Lema\^{\i}tre \cite{lema} and Eddington \cite{edd} to his own static model, Einstein fully acknowledged the merits of the non-static Fridman-type solutions rejecting outright his cosmological $\Lambda$ term as unnecessary and unjustifiable \cite{noLam}.

Interestingly, in 1931, Lema\^{\i}tre \cite{lema60} did not appreciate the disown by Einstein of his cosmological constant $\Lambda$, which, instead, was retained by the belgian cosmologist an essential ingredient of the physical Universe for a number of reasons, one of which connected also with quantum mechanics, which, however, convinced neither Einstein nor the scientific community, at least until the end of the nineties of the last century \cite{lumi}.
His \virg{hesitating} model was characterized by positive spatial curvature ($k = +1$), and a positive cosmological constant so that its perpetual expansion is first decelerated, then it enters an almost stationary state, and finally it accelerates, thus resolving the problem of the age of the Universe and the time required for the formation of galaxies.

The formal aspects of the homogeneous and isotropic expanding models were clarified and treated in a systematic, general approach in the first half of the thirties of the last century by Robertson \cite{rob29, rob33, rob35} and Walker \cite{walk35}. Today, the spacetime tensor $\tensorsym{g}$ of standard expanding cosmologies is commonly named as Fridman-Lema\^{\i}tre-Robertson-Walker (FLRW) metric (see Section \ref{flrw}).
\subsection{The Einstein-de Sitter Model}\lb{eindesmodel}

In 1932, Einstein and de Sitter \cite{eindes} published a brief note of two pages whose aim was to simplify the study of cosmology. About their work, as reported by Eddington \cite{eddane}, Einstein would have told him: \virg{I did not think the paper very important myself, but de Sitter was keen on it}, while de Sitter wrote to him: \virg{You will have seen the paper by Einstein and myself. I do not myself consider the result of much importance, but Einstein seemed to think that it was}. At any rate, such an exceedingly simplified solution, characterized by dust-like, pressureless matter, $k = 0,~\Lambda = 0$ and perpetual, decelerating expansion, served as \virg{standard model} over about six decades, to the point of curb researches on other models. In it, mutual distances among test particles grow as $t^{2/3}$. Such a behaviour is unstable in the sense that it can only occur if $k = 0$ exactly; for tiny deviations from such a value, the expansion would gradually depart from the trajectory of the Einstein-de Sitter model. Actually, it represented the best description of the cosmic expansion as it was known for the next sixty years. The fact that the observed behaviour of the physical universe was still so close to that particular expansion rate  suggested that the instability had not yet had the time to manifest itself significantly. But, after all, the universe had been expanding for about several billions of years, as if it had started just from the very spacial initial conditions of the Einstein-de Sitter model. This peculiar situation, later known as \virg{the flatness problem}, motivated, among other things, the studies on the cosmic inflation in the eighties of the last century \cite{staro, guth, linde}. The Einstein-de Sitter model has now been abandoned, also because it would imply a too short age of the Universe given by \eqi t_0 = \rp{2}{3}\rp{1}{H_0} = 9.6~\textrm{Gyr}\eqf

For a recent popular account on the panoply of possible Universes allowed by GTR, see \cite{barrowbook}.

In passing, let us note that the expanding cosmological models by GTR, along with the associated finite age of the Universe, represent the framework to correctly solve the Olbers paradox \cite{soluol1, soluol2}.
\subsection{Some Peculiar Characteristics of the FLRW Models}\lb{flrw}

The assumptions of homogeneity {\em and} isotropy of the spatial sections of the FLRW models are of crucial importance. It must be stressed that they are, in general, distinct requirements. Homogeneity does not generally imply isotropy: for instance, think about a universe filled with galaxies whose axes of rotation are all aligned along some specific spatial direction, or a wheat field where the ears grow  all in the same direction. Conversely, a space which is isotropic around a certain point, in the sense that the curvature is the same along all the directions departing from it, may well not be isotropic in other points, or, if some other points of isotropy exist, the curvature there can be different from each other: an ovoid surface is not homogenous since its curvature varies from point to point, but the space is isotropic around its two \virg{vertices}. Instead, the same vale of the curvature in all the directions, \emph{i.e}., the same amount of isotropy, around all points of space implies homogeneity \cite{weincosmo}. As far as our location is concerned, it can be said phenomenologically that isotropy about us holds in several physical aspects to a high level of accuracy, as demonstrated, e.g., by the CMB which is isotropic at a  $10^{-5}$ level. In view of the Copernican spirit, it is commonly postulated that every other observer located everywhere would see the same situation, thus assuring the homogeneity as well: it is the content of the so-called Cosmological Principle. The fact that the curvature of the spatial parts of the FLRW models is the same everywhere, and that they are expanding over time, according to the Weyl principle \cite{weyl23}, admit a peculiar foliation of the spacetime which allows for an unambiguous identification of the spatial sections of simultaneity  and of the bundle of time-like worldlines orthogonal to them as worldlines of fundamental observers at rest marking a common, cosmic time. Thus, it is possible to describe the spacetime of the Universe as the mathematical product of a $3-$dimensional Riemannian space with the temporal axis. In comoving dimensionless spatial coordinates $r,~\theta,~\phi$, the line element can be written as
\eqi (ds)^2 = c^2 (dt)^2 - S(t)\qua{\rp{(dr)^2}{1- k r^2} + r^2(d\theta)^2 + r^2\sin^2\theta(d\phi)^2}\eqf

The Einstein field equations, applied to the FLRW metric with a pressureless cosmic fluid as standard source with matter and radiation densities $\rho_{\rm m},~\rho_{\rm r}$, respectively, yield the Fridman equation
\eqi\dot S^2 = \rp{8}{3}\pi G\ton{\rho_{\rm m} + \rho_{\rm r}}S^2 - k c^2 + \rp{1}{3}\Lambda c^2 S^2\eqf By defining the Hubble parameter as \eqi H\doteq\rp{\dot S}{S}\eqf and the critical density as \eqi \rho_{\rm crit}\doteq \rp{3 H^2}{8\pi G}\eqf it is possible to recast the Fridman equation in the form \eqi\upOmega_{\rm m} + \upOmega_{\rm r} + \upOmega_{\Lambda}+\upOmega_k = 1\eqf or also \eqi\upOmega_{\rm tot}=1-\upOmega_k\eqf by posing \eqi\upOmega_{\rm tot}\doteq \upOmega_{\rm m} + \upOmega_{\rm r} + \upOmega_{\Lambda}\lb{Omegaeqz}\eqf with the dimensionless parameters entering \rfr{Omegaeqz} defined as \begin{align}
\upOmega_{\rm m}   &\doteq \rp{8}{3}\rp{\pi G\rho_{\rm m}}{H^2}>0\\ \nonumber
\upOmega_{\rm r}   &\doteq \rp{8}{3}\rp{\pi G\rho_{\rm r}}{H^2}>0 \\ \nonumber
\upOmega_{\Lambda} &\doteq \rp{1}{3}\rp{\Lambda c^2}{H^2}\lesseqqgtr 0\\ \nonumber
\upOmega_k         &\doteq -\rp{k c^2}{S^2 H^2}\lesseqqgtr 0
\end{align}
At present epoch, $\upOmega_{\rm r,0}\sim 0$, so that the normalized Fridman equation reduces to \eqi\upOmega_{\rm m,0}  + \upOmega_{\Lambda,0}+\upOmega_{k,0} = 1\eqf or also \eqi\upOmega_{{\rm tot}, 0} = 1-\upOmega_{k,0}\eqf

Results collected in the last twenty years from a variety of observational techniques (e.g., SNe Ia \cite{accel1, accel2, accel3}, Baryon acoustic oscillations \cite{BAO}, WMAP \cite{wmap}, Planck \cite{planckS}), interpreted within a FRLW framework,  point towards an observable universe whose spatial geometry is compatible with an Euclidean one (such a possibility, in view of the unavoidable error bars, is impossible to be proved experimentally with certainty: on the contrary, it could be well excluded should the ranges of values for $\upOmega_{{\rm tot},0}$ did not contain 1), and whose dynamical behaviour is characterized by a small positive cosmological constant $\Lambda$ which makes it accelerating at late times. By assuming $\upOmega_{{\rm tot}, 0}=1$ {\em exactly}, as allowed by the experimental data and predicted by the inflationary paradigm, the values for the other normalized densities are inferred by finding \cite{wmap, planckS} \eqi\upOmega_{{\rm m},0}\sim 0.3,\upOmega_{\Lambda,0}\sim 0.7\eqf
%
\section{Summary}\lb{fine}

\scalebox{0.97}{After its birth, GTR  went to fertilize and seed, {directly as well as indirectly}, many branches of disparate} { sciences as mathematics \cite{singu, hawk, soluz, soluz2, frame1, frame2}, metrology \cite{metro1, metro2, metro3, metro4},  geodesy \cite{mul08, kop011, com013}, geophysics \cite{geo1, geo2, geo3},} astronomy \cite{astro1, astro2, astro3, kop011b, kop011c, astro4}, astrophysics \cite{astrop1, astrop2, astrop3, astrop4, astrop5}, cosmology \cite{cosmo1, cosmo2, cosmo3, Gasp}, not to say about the exquisite technological spin-off \cite{tech1, tech2, sted0, tech3, tech4, tech5, tech6, tech7, tech8, tech9, tech10, tech11, tech12, tech13, tech14} due to the long-lasting efforts required to put to the test various key predictions of the theory \cite{WillLRR, exp1}.
Moreover, once some of them  have been or will be successfully tested, they have or will become precious tools for determine various parameters characterizing several natural systems, often in extreme regimes unaccessible with other means: gravitational microlensing for finding extrasolar planets, even of terrestrial size \cite{tool1, tool2}, weak and strong gravitational lensing to map otherwise undetectable matter distributions over galactic, extragalactic and cosmological\linebreak scales \cite{tool3, tool4, tool5}, frame-dragging to measure angular momenta of spinning objects like stars and\linebreak planets \cite{tool6, tool7, tool8, tool9}, gravitational waves to probe, e.g., quantum gravity effects \cite{tool10}, modified models of gravity \cite{tool11, dam05} and cosmic inflationary scenarios \cite{flop1, flop2}, to characterize tight binary systems hosting compact astrophysical objects like white dwarves, neutron stars and black holes \cite{GW1, GW2, tool12, tool13, tool14, tool15}, and to investigate extremely energetic events like, e.g., supernov{\ae} \mbox{explosions \cite{tool16}}.

However, GTR has its own limits of validity, and presents open problems \cite{open}. At certain regimes, singularities plague it \cite{singu, singu1, singu2, singu3, singu4}. Connected to this issue, there is also a major drawback of the theory of gravitation of Einstein, \emph{i.e}., its lingering inability to merge with quantum mechanics yielding a consistent theory of quantum gravity \cite{quantum1, quantum2, quantum3, quantum4, quantum5, quantum6, quantum7}. Moreover, in view of the discoveries made in the second half of the last century about the seemingly missing matter to explain the rotation curves of galaxies \cite{DM1, DM2, DM3} and the accelerated expansion of the Universe \cite{accel1, accel2, accel3}, it might be that GTR need to be modified \cite{modi1, modi2, modi3, modi4, modi5, modi6} also at astrophysical and cosmological scales in order to cope with the issue of the so-called \virg{dark} \cite{dark} components  of the matter-energy content of the Universe known as Dark Matter and Dark Energy.

We consider it appropriate to stop here with our sketchy review. Now, we give the word to the distinguished researchers who will want to contribute {to} this Special Issue by bringing us towards the latest developments of the admirable and far-reaching theory of gravitation by Einstein. {At a different level of coverage and completeness, the interested reader may also want to consult the recent two-volume book \cite{Kop3, Kop4}}.

\end{document}